# Double Etch Method for the Fabrication of Nanophotonic Devices from Van der Waals Materials

Otto Cranwell Schaeper[1,2], Lesley Spencer[1,2], Dominic Scognamiglio[1], Waleed El-Sayed[3], Benjamin Whitefield[1,2], Jake Horder[1,2], Nathan Coste[1], Paul Barclay[3,4,5], Milos Toth[1,2], Anastasiia Zalogina[1,2*], and Igor Aharonovich[1,2*]

[1] School of Mathematical and Physical Sciences, University of Technology Sydney, Ultimo, New South Wales 2007, Australia

[2] ARC Centre of Excellence for Transformative Meta-Optical Systems, University of Technology Sydney, Ultimo, New South Wales 2007, Australia

[3] Institute for Quantum Science and Technology, University of Calgary, Calgary, AB T2N 1N4, Canada

[4] Department of Physics and Astronomy, University of Calgary, Calgary, AB T2N 1N4, Canada

[5] Nanotechnology Research Centre, National Research Council of Canada, Edmonton, AB T6G 2M9, Canada

*anastasiia.zalogina@uts.edu.au; igor.aharonovich@uts.edu.au

*The integration of van der Waals (vdW) materials into photonic devices has laid out a foundation for many new quantum and optoelectronic applications. Despite tremendous progress in the nanofabrication of photonic building blocks from vdW crystals, there are still limitations, specifically with large-area devices and masking. Here, we focus on hexagonal boron nitride (hBN) as a vdW material and present a double etch method that overcomes problems associated with methods that employ metallic films and resist-based films for masking. Efficacy of the developed protocol is demonstrated by designing and fabricating a set of functional photonic components – including waveguides, ring resonators and photonic crystal cavities. The functionality of the fabricated structures is demonstrated through optical characterization over several key spectral ranges. These include the near-infrared and blue ranges, where the hBN boron vacancy ($V_B^-$) spin defects and the coherent B center quantum emitters emit, respectively. The double etch method enables fabrication of high-quality factor optical cavities and constitutes a promising pathway toward on-chip integration of vdW materials.*

Integrated quantum photonics has attracted significant attention by introducing photons as qubits that are robust against decoherence, and by using resonant structures for photon manipulation[1-6]. Confinement and control of light at the nanoscale require advanced protocols for the nanofabrication of photonic building blocks, including waveguides, cavities and resonators[5]. In recent years, van der Waals (vdW) crystals have emerged as a promising platform for quantum photonic devices [7-11]. For example, transition metal dichalcogenides offer a combination of nonlinear effects and a high refractive index, which is advantageous for light confinement and photon guiding[12].

Hexagonal boron nitride (hBN) is particularly attractive as a vdW material for quantum applications[13]. Due to its wide bandgap of ~6 eV, it can host a variety of ultra-bright optically active defects that operate at room temperature and are suitable for quantum information processing[14-19]. In addition, hBN also hosts ensembles of spin defects that are of interest for quantum sensing[20-24]. Some of these defects, can be engineered deterministically, with reproducible emission wavelengths which is crucial for scalability of on-chip devices[15]. A key challenge in utilizing these defects in practical on-chip devices is coupling them to the modes of nanostructures used to generate, manipulate and collect the emitted photons. These nanostructures include waveguides, ring resonators, nanobeam photonic crystal cavities, metasurfaces or topological photonic structures[3, 5]. One approach is to stack vdW monolayers or multilayers with photonic devices fabricated from other materials (e.g. silicon nitride or titanium oxide), relying on coupling via evanescent fields [9, 25, 26]. However, the efficiency of this so-called hybrid approach is limited due to suboptimal overlap between the emitter and the field of an optical mode. An alternative is the monolithic approach, where emitters and photonic nanostructures are integrated into the same material [27-29]. This can result in maximum coupling efficiency, minimal losses, and high-quality factors.

In the current work, we present a fabrication method tailored for vdW materials, employing a double etch approach for mask transfer using an ion beam etcher (IBE) and an inductively coupled plasma reactive ion etcher (ICP RIE). This method addresses limitations of existing vdW material nanofabrication protocols that employ metallic films and resist films for masking (more information about these limitations is provided in the Supplementary Information).

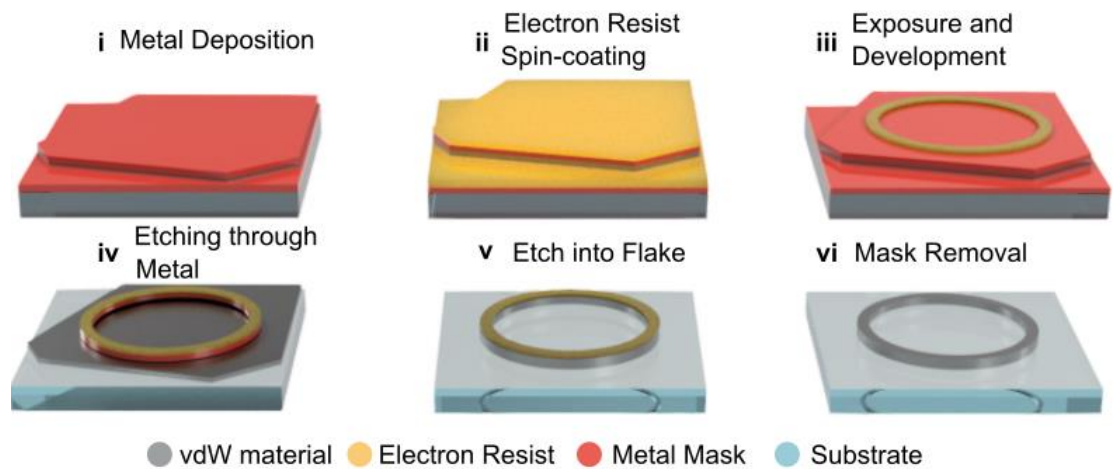

*Figure 1. Schematic of the double etch protocol. (i) Metal deposition onto an exfoliated flake of hBN. (ii) Electron-beam resist spin-coating. (iii) Resist exposure and development. (iv) First etching step through the exposed metal mask using IBE. (v) Second etching step through vdW material using ICP RIE. (vi) Metal mask and resist removal.*

The developed double etch method is based on the complementary etching chemistries used to transfer patterns from an electron-beam resist into tungsten via a fluorine etch. This method eliminates the need for sonication and lift-off, making it compatible with vdW materials. This is crucial, because unlike nanofabrication of 3D materials, vdW crystals can easily get lost during the sonication process by disconnecting from the substrate. The double etch fabrication protocol is illustrated schematically in Figure 1. First, a vdW flake is coated homogeneously with a layer of metal (i), followed by spin-coating of an electron-beam resist (ii). A pattern is then defined in the resist, where the unexposed resist becomes the region protected by the metal

(iii). The resist layer now acts as a hard mask for etching the metal using ICP RIE (iv), eliminating the risk associated with the lift-off process, which often leads to removal and loss of the flake. Given the physical nature of IBE, the resist layer must be sufficiently thick and resistant to IBE etching. In this particular case, a 300-nm layer of polymethyl methacrylate (PMMA) was employed. Next, the vdW material is etched using ICP RIE through the created double mask of resist and metal (v). Finally, the double mask is removed using a mixture of sulfuric acid and hydrogen peroxide (piranha solution) (vi). See the methods section and the SI for more details. Note that conventionally, a hard mask is used with an intermediate adhesive layer between the resist and metal layers. Here, the metal hard mask coats the entire sample uniformly, eliminating the need for an adhesive layer and simplifying both the metal deposition and removal steps.

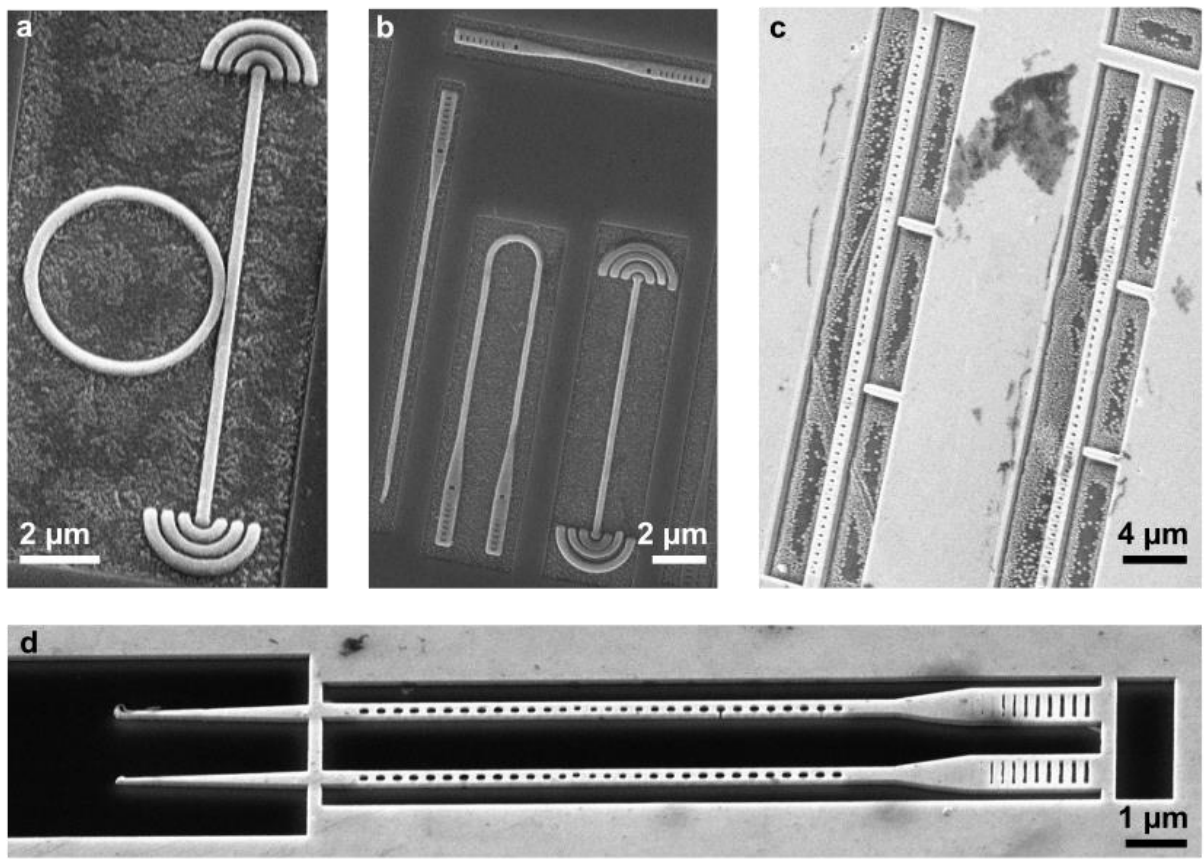

*Figure 2. Scanning electron microscope images of photonic devices fabricated from hBN using the double etch method. (a) Microring resonator coupled to a bus waveguide. (b) Four different waveguide geometries, including one with a bend geometry. (c) One-dimensional (1D) photonic crystal cavities (PCC) held by two spokes, suitable for optomechanics. (d) 1D PCC waveguides with tapered ends and grating couplers.*

To demonstrate the utility of the double etch method, we designed and fabricated photonic components with various architectures hosting emitters at different emission wavelengths. For clarity and usefulness, we focus on two regions of interest in hBN: (1) the near-infrared region, where the spin-active boron vacancy ($V_B^-$) defects emit and (2) the blue spectral range where the B centers with a zero phonon line (ZPL) at 436 nm emit. The B centers are currently the most promising for deployment in quantum information protocols due to their coherent properties, hence we fabricated devices in this spectral range[14, 30, 31]. Lastly, we also fabricate a one-dimensional (1D) photonic crystal cavity (PCC) with a resonance in the infrared range, which has not been realized previously using hBN. These functional nanostructures are foundational to future hBN quantum circuitry and offer opportunities for scalable, reproducible, and robust on-chip devices, overcoming challenges with existing vdW material fabrication methods.

Figure 2 shows scanning electron microscope (SEM) images of these devices. All structures are based on hBN material exfoliated onto a Si substrate with a dissipative $SiO_2$ layer. Different thicknesses of hBN flakes were chosen to maximize light confinement within the designed photonic structures. The microring resonator in Figure 2 (a) is designed to host $V_B^-$ defects emitting at 760 nm (1.6 eV). The microring resonator, with a diameter of 5 μm, is coupled to a bus waveguide with couplers via a 100-nm gap, which allows for the collection of emission from the microring. Such structures are designed for on-chip distributed quantum sensing applications discussed further in Figure 3.

A set of different waveguides for guiding single photons, featuring variously shaped Bragg grating couplers for efficient excitation and collection, are shown in Figure 2 (b). These waveguides are designed to contain B centers with an emission wavelength of 436 nm (2.8 eV) created using electron beam irradiation. Each waveguide has dimensions of 200 nm in width and thickness, ensuring good confinement for light guiding.

Figure 2 (c) shows an example of a 1D PCC. This particular cavity was undercut using a potassium hydroxide solution (KOH), creating an air gap and eliminating the need for the $SiO_2$ layer. These undercut cavities were fabricated to function as optomechanical systems and are suitable for studying thermally driven mechanical resonance [32].Finally, one of the most critical nanophotonic structures - a tapered waveguide with a 1D PCC and a Bragg grating coupler, is demonstrated in Figure 2 (d) [3, 33-35]. These structures are often employed for highly efficient side collection with a matching tapered single mode fiber.

To characterize the functionalities of the fabricated photonic devices, we measured a microring resonator containing the $V_B^-$ spin defects. Two different excitation configurations were used when measuring this structure, as shown in the schematics in Figure 3 (a, b). In the first configuration we excited the devices through one of the couplers and collected through the other coupler. In the second configuration we directly excited the $V_B^-$ spin defects on the ring and collected through the coupler. Photoluminescence (PL) measurements were performed using a lab-built scanning PL setup featuring decoupled excitation and collection. The device was excited using a 532 continuous wave (CW) laser through a 100X 0.9 NA objective and was collected through the same objective.

The spectra in Figure 3(c, d) conclusively show that the devices operate as intended. The $V_B^-$ defects exhibit the broad emission centered at around 760 nm. Whispering gallery modes (WGMs) are clearly visible and evident as sharp peaks within the spectrum. The quality factor of the WGMs is defined here as $Q = \lambda/\Delta\lambda$ (where $\lambda$ is a resonant wavelength, and $\Delta\lambda$ is a full width at half maximum). The measured $Q$ is ~300 at 684 nm for the current set of devices. The $Q$ of the resonators can be further improved by undercutting the structures or fabricating larger resonators.

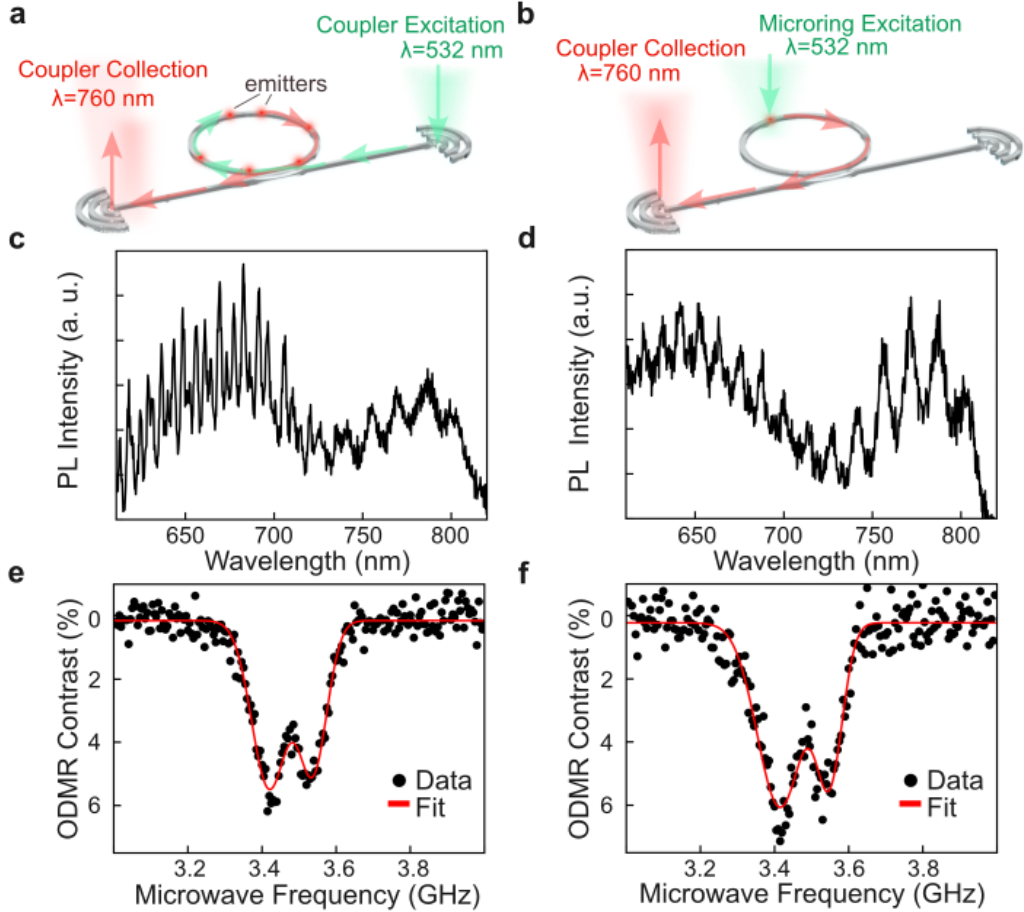

*Figure 3. On-chip distributed quantum sensing.* Schematics of two excitation schemes of the $V_B^-$ defects: excitation through the bus waveguide (a), and direct excitation from the microring (b). Collection in both cases is done using the second coupler. (c, d) Photoluminescence (PL) intensity spectra, collected using both configurations (excitation through the bus waveguide and excitation from the ring). The broad emission around ~760 nm corresponds to the $V_B^-$ ensemble, and the sharp peaks are the WGMs. (e, f) Optically detected magnetic resonance (ODMR) of the $V_B^-$ spin defects measured at zero magnetic field under the same configurations.

With this device, we further demonstrate on-chip distributed quantum sensing measurement via optically detected magnetic resonance (ODMR) of the $V_B^-$ defects embedded in the microring. Using the same setup configurations as mentioned above (Figure 3 (a, b)), we collected the ODMR signal by measuring the emission intensity difference as the microwave signal is swept across the $V_B^-$ ground state (see Methods). The double dip at ~3.4 and 3.55 GHz corresponds to the splitting of the ground state of the $V_B^-$ ensemble and serves as a signature for magnetic field sensing[20]. An ODMR contrast of ~6% was obtained for both configurations (Figures (e, f)), which shows that there is no degradation in OMDR contrast within the device. Importantly, the $V_B^-$ defects are excited equally well both from the top of the microring, as well as through the bus waveguide. The latter offers a unique feature whereby the devices can probe magnetic fields in a delocalized geometry, where the excitation and collection is performed through couplers and bus waveguides [36]. Such an approach enables distributed quantum sensing across a larger area of the chip employing several large coupled microring resonators, which will become feasible.

We now turn to the waveguides designed to host emitters in the blue spectral range (B centers), with a ZPL at ~436 nm. After fabrication, B centers were deterministically placed by electron beam irradiation and resonant excitation (photoluminescence excitation (PLE)) was performed. Observing a narrow linewidth from resonant excitation is important for future Hong-Ou-Mandel experiments as well as advanced quantum optics measurements [14, 37]. The device is shown in Figure 4 (a): the emitter is resonantly excited (light blue) and its emission is collected (dark blue) through the same coupler. A mirror at the opposite end of the waveguide allows for increased collection efficiency. This measurement is carried out at 5 K. We measure the emission from the phonon sideband as a function of the laser detuning to the center of the ZPL. The results are displayed as the black circles in Figure 4 (b). The same experiment was repeated by exciting and collecting directly on top of the emitter (blue circles in Figure 4 (b)). The excitation and collection of the B center through the coupler yielded a resonant linewidth of ~7.7 GHz, whereas excitation and collection on top of the emitter resulted in a slightly broader line of ~8.5 GHz. The inset of Figure 4 (b) shows a spectrum of waveguide emission from the coupler under off-resonant (405 nm) excitation, showing the ZPL of the emitter (at room temperature).

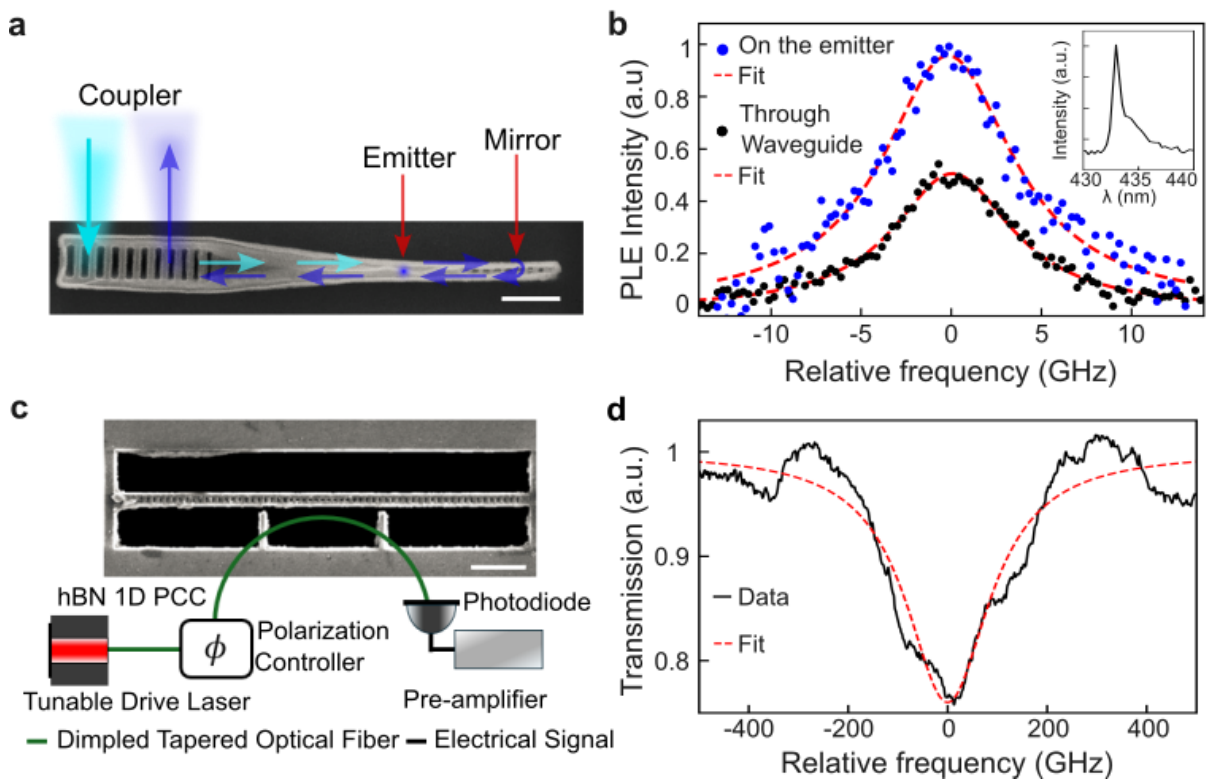

*Figure 4. Optical characterization of fabricated devices. (a) SEM image with the schematic of excitation (light blue) and the collection (blue) of B centres. Scale bar is 1 μm. (b) Resonant photoluminescence excitation (PLE) spectra for the emitters excited through the waveguide (filled black circles) and excited and collected from the same spot (filled blue circles) collected*

*at 5K. Inset: Photoluminescence (PL) spectrum collected at room temperature. (c) Schematic of the optical setup used to measure the 1D PCC with SEM of 1D PCC image at the top. Scale bar is 5 μm. (d) Transmission spectrum. The estimated Q factor from this spectrum is ~1000.*

Finally, we test the 1D PCC at the telecom range. This range remains unexplored since there have not been found emitters in hBN. We demonstrate the first evidence of a high $Q$ cavity at ~1600 nm. Figure 4 (c) shows the schematic of the setup used to measure this device. Laser light is sent into a dimpled tapered optical fiber, which evanescently couples into the beam cavity when positioned in a close proximity. The laser frequency is then scanned from 1500 to 1700 nm. The transmission spectrum is measured using a photodetector. For the fabricated device (SEM image at the top of Figure 4 (c)), we obtained an intrinsic $Q$ of ~1000 at the wavelength of ~1600 nm, as shown in Figure 4 (d).

To conclude, we introduced a double-etching method to fabricate scalable photonic devices from vdW materials and demonstrated it with hBN. We fabricated cavities and microring resonators with bus waveguides, that are suitable to be deployed for quantum circuitry. A key advantage of this method is elimination of the need of lift-off or sonication during the fabrication process. Using the double etching method, we realized both on-chip distributed quantum sensing with $V_B^-$ defects as well as coherent excitation of the B centers. Furthermore, we showed that the method is suitable to engineer high-$Q$ cavities for the telecom range that had not been realized to date. Such fundamental advances in fabrication could be applicable to other vdW materials, particularly TMDCs, thereby opening up possibilities for compact and efficient quantum photonic devices.

## Methods

**Fabrication.** Pristine hBN flakes sourced from the National Institute for Materials Science (NIMS) were mechanically exfoliated using scotch tape and then transferred using Gel-Pak strips onto 285-nm $SiO_2$ on Si substrates. These samples were annealed in a lindberg blue mini-mite tube furnace in an ambient atmosphere at temperatures of 500°C, 650°C, and 850°C to create B centers in hBN as well as increase the adhesion of the hBN flakes to the sample surface and remove the polymer contamination. Exfoliated hBN flakes for the set of waveguides in Figure 2 (b) were annealed on a hot plate instead, at 500°C, to remove the polymer contamination. Next, hBN flakes were preselected based on the thickness requirements for fabrication using an optical microscope. A 25-nm layer of nickel is then deposited across the substrate and hBN flakes using an electron beam evaporator (AJA ATC-1800E). A layer of electron-beam resist polymethyl methacrylate (PMMA) A5 (AllResist gmbh) was spin-coated on the sample at 5000 rpm for 30 s resulting in a layer of polymer ~300 nm thick, the sample was then baked at 180°C for three minutes. The polymer was then patterned using electron beam lithography (Elionix ELS-F125) at 125 kV and 1 nA. Next, the exposed pattern was developed in 1:3 IPA:MIBK for 60 s, rinsed in IPA for 20 s, and gently dried with a nitrogen gun. Following the development, the pattern is transferred by the use of an ion beam etcher (Intlvac) at 6 sccm Ar, 1.5 A of discharge, 0.150 A of emission to the LFN, 12 sccm of Ar to the ion source, 400 V beam voltage, 150 mA beam current, 60 V accelerator voltage, with a stage temperature of 20°C for 90 s. The nickel and any remaining resist then act as the hard mask when the mask is transferred into the hBN using a reactive ion etching in ICP-RIE system (Trion) at 1 mT, 5 sccm $SF_6$, 60 sccm Ar, 300W RF and 1W ICP with the etch rate of ~5 nm/s. Then the nickel mask is removed using piranha solution (3:1 sulphuric acid: hydrogen peroxide) at room temperature. Once the nickel mask had been removed, a dual beam system was used to implant both $V_B^-$ and B center emitters (Thermo Fisher Scientific Helios G4). $V_B^-$ centers were implanted in the microring (Figure 2 (a)), using a 30 keV nitrogen ion beam to irradiate an area with a fluence of $1\times10^{14}$ ions/cm$^2$. B centers were implanted using a 3 keV, 1.6 nA electron beam with a 5 ms dwell time. The dwell time was varied over multiple waveguides to achieve an electron dose of $7.5\times10^{10}$ to $3.0\times10^{11}$.

**Photoluminescence Measurements.** $V_B^-$ emission was measured using a lab-built confocal microscopy setup with a fixed 532 CW laser excitation. A 4F system and scanning mirror allowed for decoupling of the collection from the excitation. A 100X 0.9 NA objective (Nikon) was used for both excitation and collection, directing light to a spectrometer (Princeton Instruments Acton SP2300) or an avalanche photodiode (APD) (Excelitas SPCM AQRH-14-FC) through a multimode fiber. B center emission was measured using a lab-built confocal microscopy setup with 0.82 NA objective lens (Attocube). Resonant excitation was performed with a frequency doubled Ti:sapphire scanning laser (M Squared) scanned at a rate of 1 GHz/s. Photons from the phonon sideband were collected with a 442 nm long-pass filter and directed to a spectrometer (Andor) or an APD through a multimode fiber.

**Optically Detected Magnetic Resonance (ODMR).** Continuous wave ODMR was performed on the $V_B^-$ ensemble by sweeping a radio frequency (RF) signal from 3 to 4 GHz. The RF signal was supplied by a copper wire antenna suspended above the sample. At frequencies of 3.4 and 3.55 GHz, the electron spin is driven between the $m_s = 0$ to $-1$ and the $m_s = 0$ to $+1$ respectively. The electron is then excited through 532 nm optical pumping resulting in a spin dependent decay and leading to a decrease in photon counts when the spin transitions are driven. The frequency sweep alternated between RF on and off for signal and reference measurements. The photon counts for each frequency were collected by an avalanche photodiode (APD) and the ODMR contrast percentage was then calculated using the equation: Contrast (%) = (Signal-Reference)/Reference * 100. All ODMR measurements were completed at zero external magnetic field and fitted with a double Gaussian curve.

**Acknowledgments**
We thank Angus Gale and Karin Yamamura for assistance with the ion/electron irradiation for emitter creation. We thank Milad Nonahal and Giorge Gemisis for all the advice they provided. The authors acknowledge financial support from the Australian Research Council (CE200100010, FT220100053) and the Office of Naval Research Global (N62909-22-1-2028). The authors acknowledge Takashi Taniguchi (the National Institute for Materials Science) for providing hBN crystal. The authors acknowledge the use of the fabrication facilities as well as scientific and technical assistance from the Research and Prototype Foundry Core Research

Facility at the University of Sydney, being a part of the the NCRIS-enabled Australian National Fabrication Facility (ANFF), and the UTS facilities, being a part of the ANFF-NSW node.